\documentclass[a4paper,11pt]{article}
\pdfoutput=1 

\usepackage{jinstpub} 

\graphicspath{{figures/}} 
\title{\boldmath Investigation of energy spectrum and chemical composition of primary cosmic rays in 1--100 PeV energy range with a UAV-borne installation}

\author[a,1]{D.~Chernov,\note{Corresponding author.}}
\author[a]{E.~Bonvech,}
\author[c,d]{M.~Finger~Jr.,}
\author[c,d]{M.~Finger,}
\author[b]{V.~Galkin,}
\author[b]{V.~Ivanov,}
\author[a,b]{D.~Podgrudkov,}
\author[a]{T.~Roganova}
\author[a,b]{and I.~Vaiman}

\affiliation[a]{Lomonosov Moscow State University, Skobeltsyn Institute for Nuclear Physics, Moscow, Russian Federation}
\affiliation[b]{Lomonosov Moscow State University, Faculty of Physics, Moscow, Russian Federation}
\affiliation[c]{Charles University, Faculty of Mathematics and Physics, Prague, Czech Republic}
\affiliation[d]{Joint Institute for Nuclear Research, Dubna, Russian Federation}

\emailAdd{chr@dec1.sinp.msu.ru}

\abstract{A new project is developed with the implementation of a relatively new method of studying the primary cosmic ray --- the registration of extensive air showers' optical Vavilov-Cherenkov radiation (Cherenkov light) reflected from the snow surface. The aim of the project is the study of the cosmic ray mass composition in the energy range of 1--100 PeV by detecting the reflected extensive air showers' Cherenkov light. Silicon photomultipliers are planned to be used as the main photosensitive element of the detector and an unmanned aerial vehicle will is planned to lift the measuring equipment over the snow-covered ground.}


\proceeding{Instrumentation for Colliding Beam Physics, INSTR20\\
from 24 to 28 February, 2020\\
Novosibirsk, Russia}

\begin{document}
\maketitle
\flushbottom

\section{Introduction}
\label{sec:intro}

The 1--100 PeV energy range is considered to be a transitional area from galactic to extragalactic cosmic rays. More than 50 years ago a change in the slope of the energy spectrum of primary cosmic rays (PCR) was detected at around 3 PeV (the ``knee''), but even nowadays new features in the structure of the spectrum are being discovered. In this regard understanding the cause of these slope irregularities is of a prime interest. One of the mechanisms behind these irregularities may be a change in the mass composition of the PCR near this region. Presently used methods allow to estimate either the average mass of PCR particles or to divide them into ``light''' and ``heavy'' groups. Most of the primary particle mass reconstruction methods are based on estimation of the depth of extensive air shower (EAS) maximum from the registered data. The relation between EAS maximum depth, zenith angle, energy and mass is derived solely from the modelling. The present day models of nucleon-nucleon interactions at high energies show some difference in the results so that the methodological uncertainties become larger than the predicted difference between nuclei mass groups.

This project is aimed at development of a unique detector using modern photodetectors based on silicon photomultipliers (SiPM), which is installed on an unmanned aerial vehicle (UAV). Currently, there are no other devices and installations that would successfully use reflected Cherenkov light (CL) registration method. The method allows to achieve the highest accuracy of estimation of the chemical composition of PCR in the analysis of the individual EAS events in comparison with existing ground installations. The successful implementation of the project will allow obtaining experimental data for the reconstruction of partial spectra for several mass groups of PCR particles (protons, helium, CNO and Fe groups) in the 1--100 PeV energy range.

\section{Method overview}

We propose to apply a reflected Cherenkov light registration method~\cite{Ant15a} for this experiment. The main idea of this method is to register the EAS CL reflected from the snowed earth surface using the compact apparatus lifted above ground. The initial idea was first introduced by A.E.~Chudakov in~\cite{Chu74}). Later the technique was successfully implemented in our earlier experiments~\cite{Ant15a, Ant15c}.

The properties of the snow surface play an important role when using the method of reflected CL registration. The results of the show optical properties studies have been repeatedly published by several groups~\cite{qun83, dum10,gre94,war82,hud06}. The simulation results show that in the wavelength range from 300 to 600~nm, the relative reflectance for pure snow is stable within 3\% for light incidence angles from 0$^\circ$ to 80$^\circ$. From the above mentioned results and the known CL spectral characteristics it can be concluded that the snow surface reflects the CL with minor spectral distortions up to 80$^\circ$ incidence angle and can be used as a ``screen'' for the EAS CL registration.

\subsection{Method advantages}

The method of reflected CL registration has a number of advantages over traditional EAS registration methods. They are as follows:
\begin{itemize}
\item The method provides a significant area of CL registration using a compact device;
\item Accurate estimation of PCR energy in an individual event in comparison with other methods thanks to the quasi-colorimetric method of energy registration;
\item The field of view of the individual sensitive elements of the device covers a significant part of the surveyed area, which allows observation the EAS CL near the shower axis, usually inaccessible to ground-based CL detector arrays. This circumstance significantly increases the accuracy of the primary particle type estimation;
\item Allows measurement of the same PCR energy range with different resolution (distance between the centres of the fields of view of neighbouring sensing elements) using variation of the detector elevation, which allows you to control the magnitude of systematic errors;
\item The small size of the required detector allows to combine the calibration techniques accessible currently only to imaging air Cherenkov telescopes (direct calibration) with large scale measurements that are accessible for only conventional ground arrays;
\item Precise timing and high level of synchronization for better primary particle arrival direction reconstruction that has a direct impact on the precision of the EAS energy estimation.
\item The compact and tight location of the detector electronics and sensitive element allows the use of the  complex local topological trigger conditions that can greatly decrease the random coincidence thus allowing to lower the measurements energy threshold.
\end{itemize}

\subsection{SPHERE-2. A previous detector realization}

\begin{figure}[t]
\centering
\includegraphics[width=1.0\textwidth]{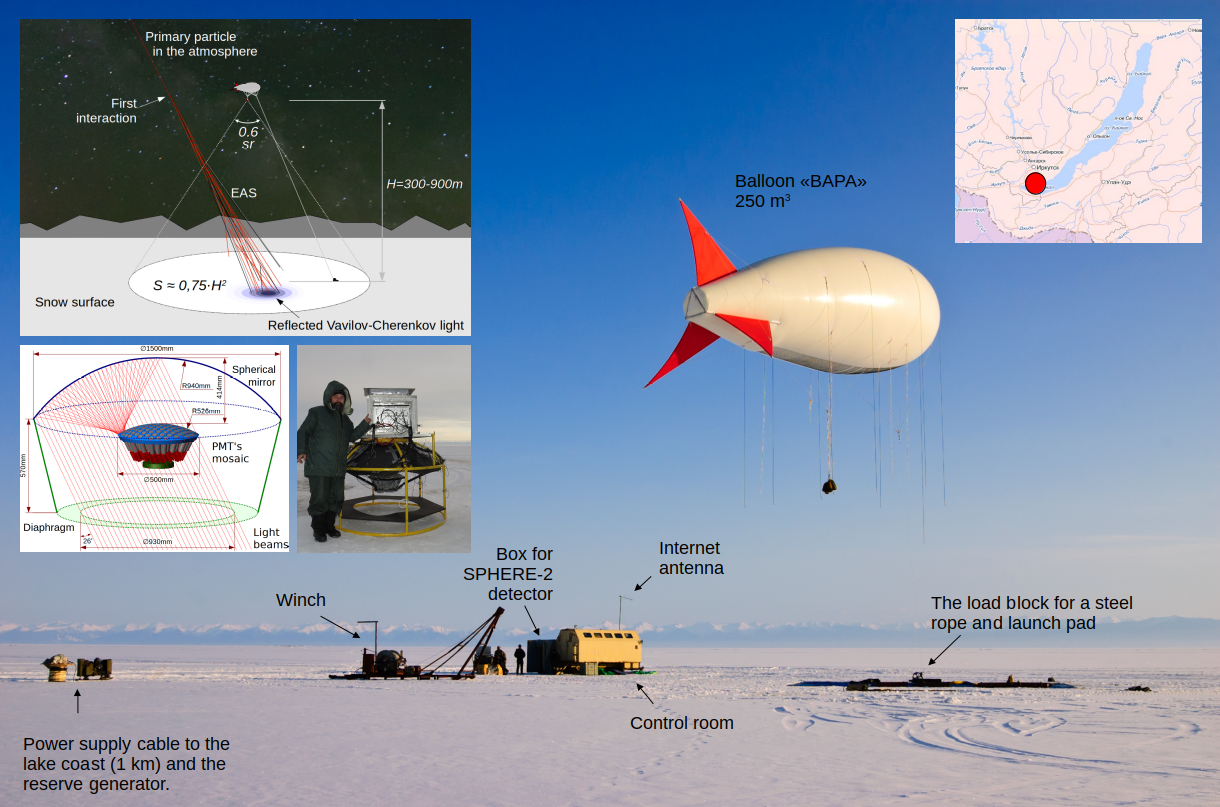}
\caption{\label{fig:Sphere_Baikal} Experiment with the SPHERE-2 installation on lake Baikal.}
\end{figure}

In the period from 2008 to 2013, a series of measurements of reflected Cherenkov light was carried out using the SPHERE-2~\cite{Ant15a,Ant15c,Ant19} balloon detector. 

The SPHERE-2 apparatus was constructed with a 1.5~m diameter spherical mirror with the 0.93~m Shmidt diaphragm window. Cherenkov light was detected by 109 PMT retina located near the focus surface of the spherical mirror. The scheme of the detector optical part is shown on the left in fig.\ref{fig:Sphere_Baikal}. The full description of the SPHERE-2 detector can be found in~\cite{Ant15a,Ant20}.

Measurements were made over the snow-covered ice of Baikal lake. The fig.\ref{fig:Sphere_Baikal} shows the landing point on the Baikal lake surface and the special BAPA balloon designed for the experiment. 

\begin{figure}[t]
\centering 
\hfill
\includegraphics[height=.25\textheight]{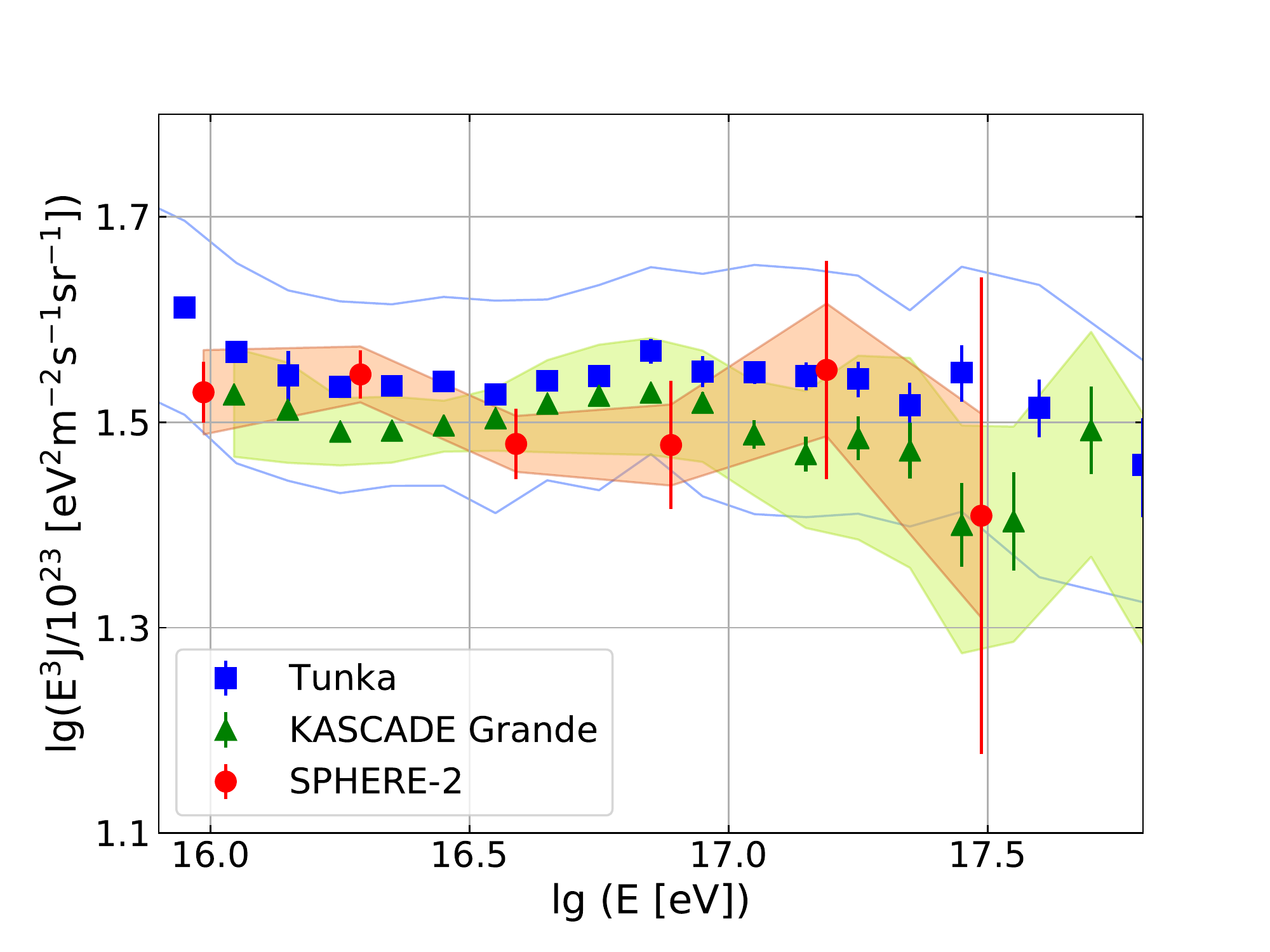}
\hfill
\includegraphics[height=.25\textheight]{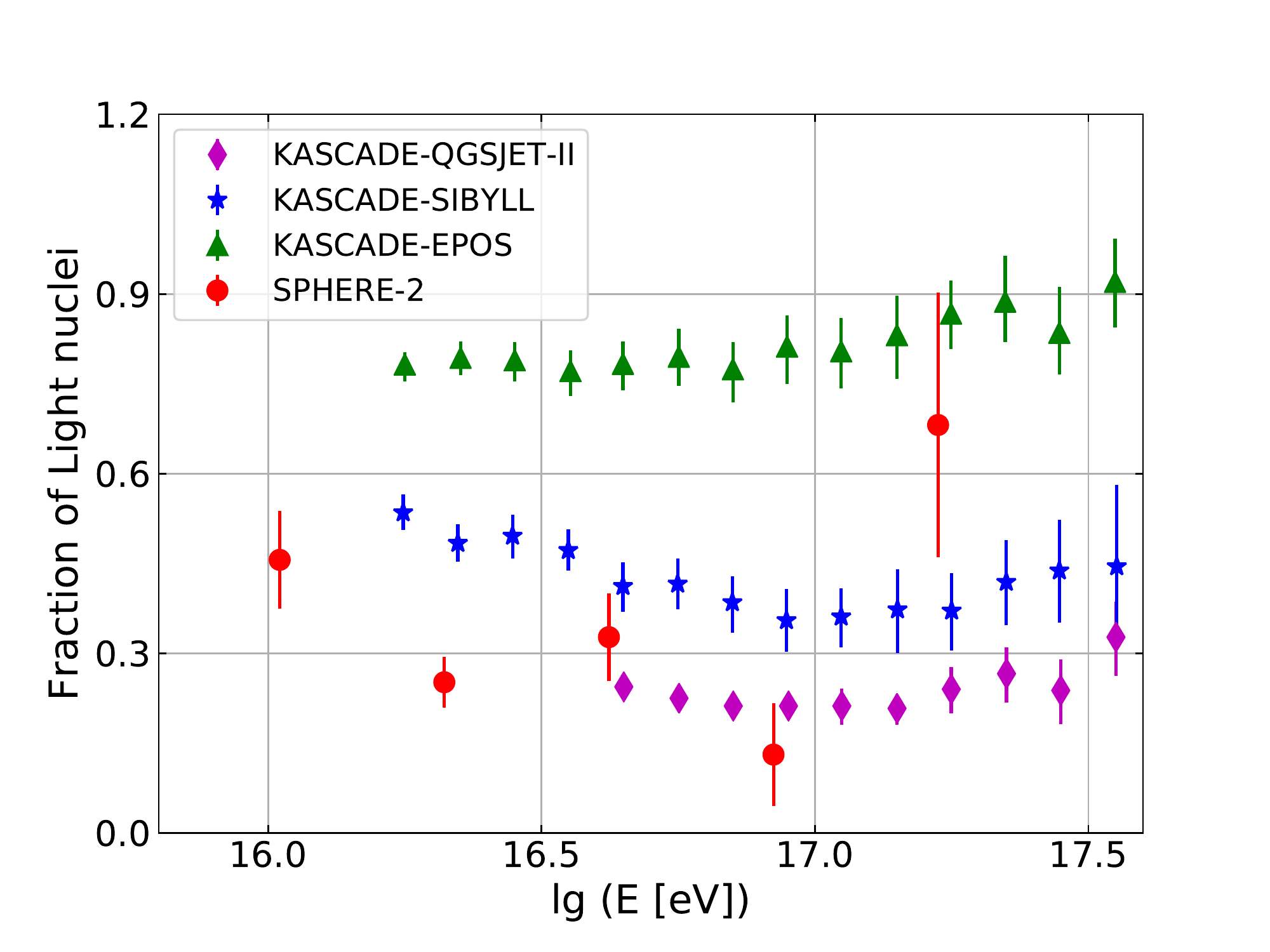}
\hfill
\caption{Results of the SPHERE-2 experiment. On the left, the energy spectrum of the PCR (red circles) is shown in comparison with the results of the Tunka-133~\cite{Tunka2020} (blue squares) and KASCADE Grande~\cite{Ape12} (green triangles) experiments. On the right the chemical composition estimates are shown (red circles) with the results of the KASCADE Grande experiment~\cite{Ape13}}
\label{fig:Sphere_results}
\end{figure}

The energy spectrum of all particles was measured and published~\cite{Ant15c}. The energy spectrum detected by the SPHERE-2 detector is shown on the fig.\ref{fig:Sphere_results}~(left) in comparison with some other ground based arrays. The systematical error of the SPHERE experiment is compatible with the KASCADE Grande one. The Tunka experiment~\cite{Tunka2020} published it probably systematical error. The light nuclei part estimation is shown on the fig.\ref{fig:Sphere_results}~(right). The actual goal is to increase the statistics of the data obtained with our technique.

The SPHERE-2 apparatus hadn't fully utilized all the possibilities of the method.
This was due to two main reasons. 
The first reason was the low sensitivity of the PMT matrix. The FEU-84 PMT~\cite{FEU84} had low quantum efficiency and covered only 30\% of the total surface leaving 70\% not sensitive to light. This reduces the expected statistical material by more than 5~times due to the increase in the registration energy threshold. The second reason was the technical difficulties associated with the need to maintain the balloon in working condition for the multiple launches of the detector. Each 10-day measurement session required at least 6 tons of cargo with helium cylinders to be transported to the measurement site, so no more than one session was possible per year. While in winter it is possible to preform up to 5~sessions per year.

Taking into account all the above mentioned difficulties we have developed a new detector based on SiPM light sensor.
Small mass of the detector will allow to abandon the cumbersome and time-consuming in use balloon and switch to the UAV as a carrier.
This experimental setup has never been used before in the study of ultra high energy PCR.
Currently, in the field of ultrahigh energy cosmic rays astrophysics, UAVs are used only for solving auxiliary tasks such as monitoring the atmosphere and calibrating ground detectors.

\section{The detector}

Good methodological accuracy of the measurements is required for the development of new experiments. 
Based on the operating experience of the SPHERE-2 detector it is possible to design a new detector that will be superior in its capabilities to its predecessor.

Advantages of the described above technique and progress in the field of microelectronics already allows to design the compact detector of the reflected EAS CL with big effective area of registration, a wide viewing angle (for PCR) and high spatial resolution. 

In comparison to a ground detector array (with an effective area of about 1 km$^2$ and more, service infrastructure etc.) a compact detector weighing up to 10~kg with similar characteristics (geometric factor) will cost less in terms of the amount of material and labor with comparable scientific results.

A compact detector that will have the following characteristics is currently in design:

\begin{itemize}
\item Sensitive area of optics (aperture input window) around 0,1m$^2$;
\item Mirror diameter 0.8 or 1~m;
\item Optical system viewing angle up to $\pm$25~degrees;
\item Number of mosaic elements 133--259 SiPM;
\item The mass of the detector less 10 kg;
\item The flight height of the detector 300--700~m;
\item Expected number of events EAS (with $E_0$ = 1--100 PeV) up to 3\,000 for season.
\end{itemize}

\begin{figure}[bt]
\centering 
\includegraphics[width=.32\textwidth,clip]{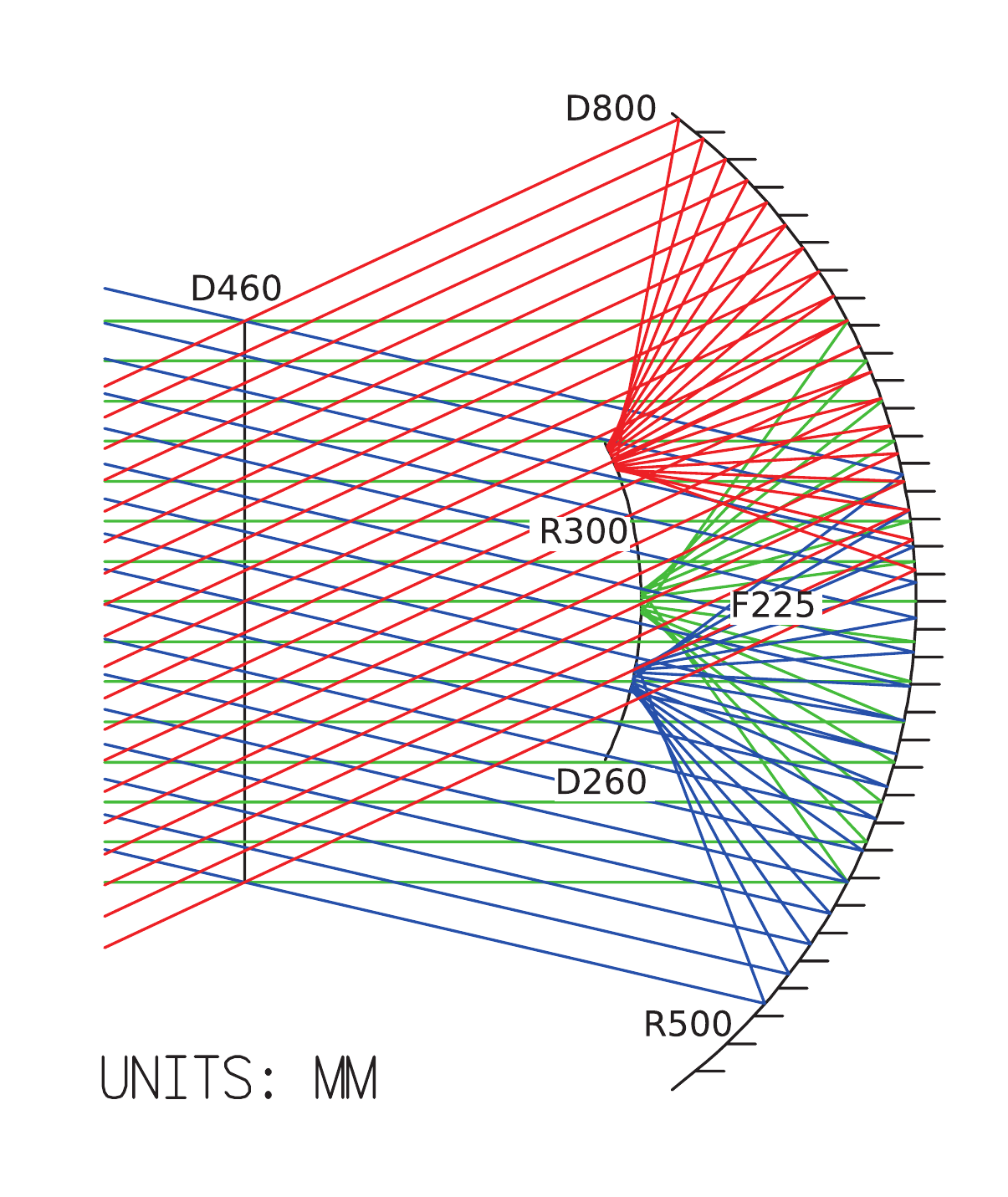}
\qquad
\includegraphics[width=.55\textwidth]{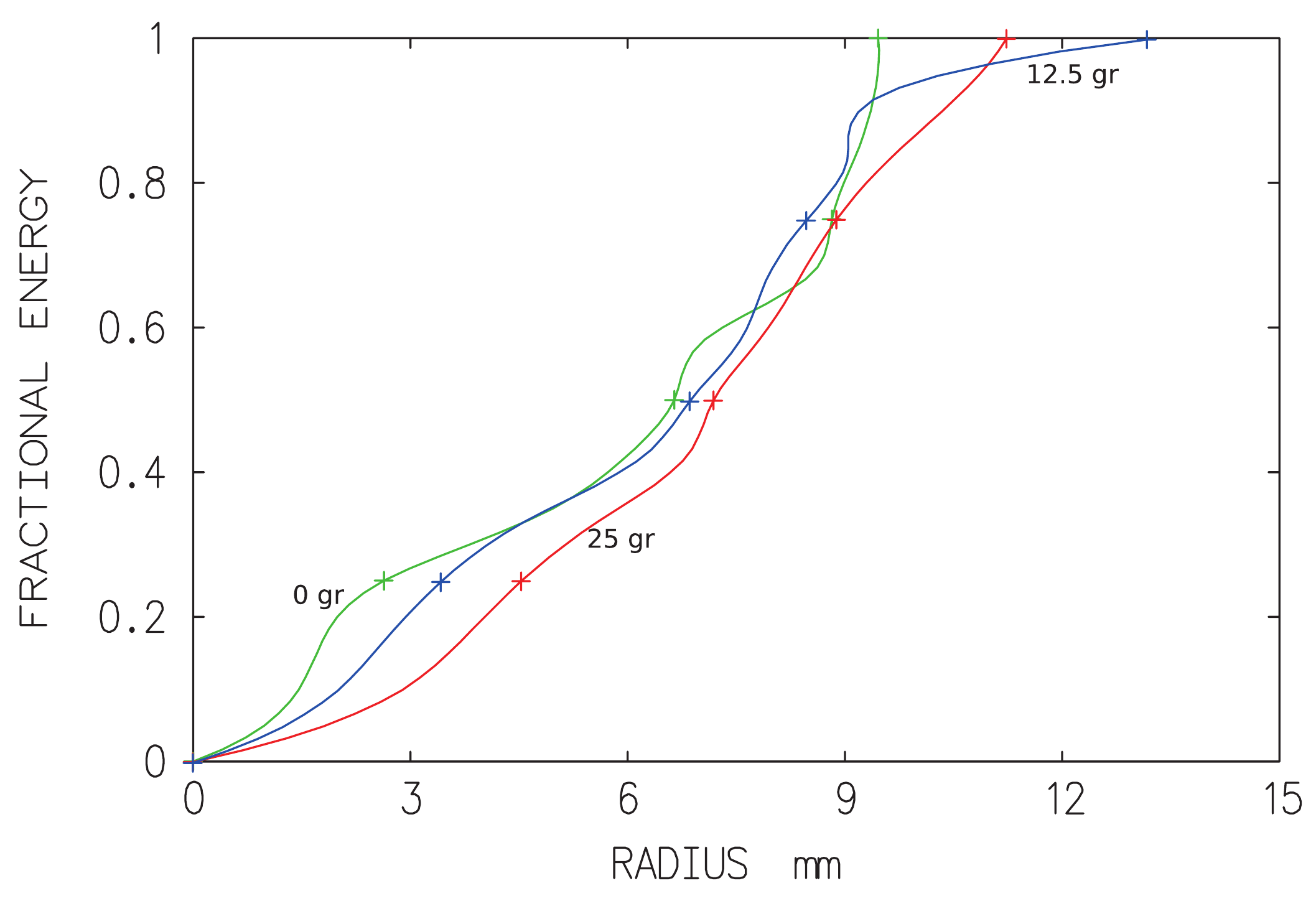}
\caption{Optical scheme of the developed detector (without shadow from the mosaic) and the fractional light energy distribution on focal plane taking into account the shadow from the mosaic. Red color corresponds to the rays coming at an angle of 25$^\circ$ from the optical axis, blue --- 12.5$^\circ$ and green --- paraxial.}
\label{fig:optic_sphere3}
\end{figure}

One of the optical scheme variants based on the simplified Schmidt scheme (no corrector plate) is shown in the left fig.\ref{fig:optic_sphere3} (the calculations were performed using OSLO EDU program~\cite{Oslo}). The rays are shown without the absorption on the backside of the SiPM mosaic for illustration only. The detector is planned to have a spherical 800~mm diameter mirror with 500~mm curvature radius. The 460~mm diameter diaphragm situated 550~mm from the mirror centre.

The image of the reflected shower is formed on the photosensitive spherical surface with 260~mm diameter and 300~mm curvature radius. The optimum distance for this set of parameters is 225~mm from the mirror. Maximum observation angle is $\pm$25$^\circ$.

\begin{figure}[bt]
\centering
\includegraphics[width=0.7\textwidth]{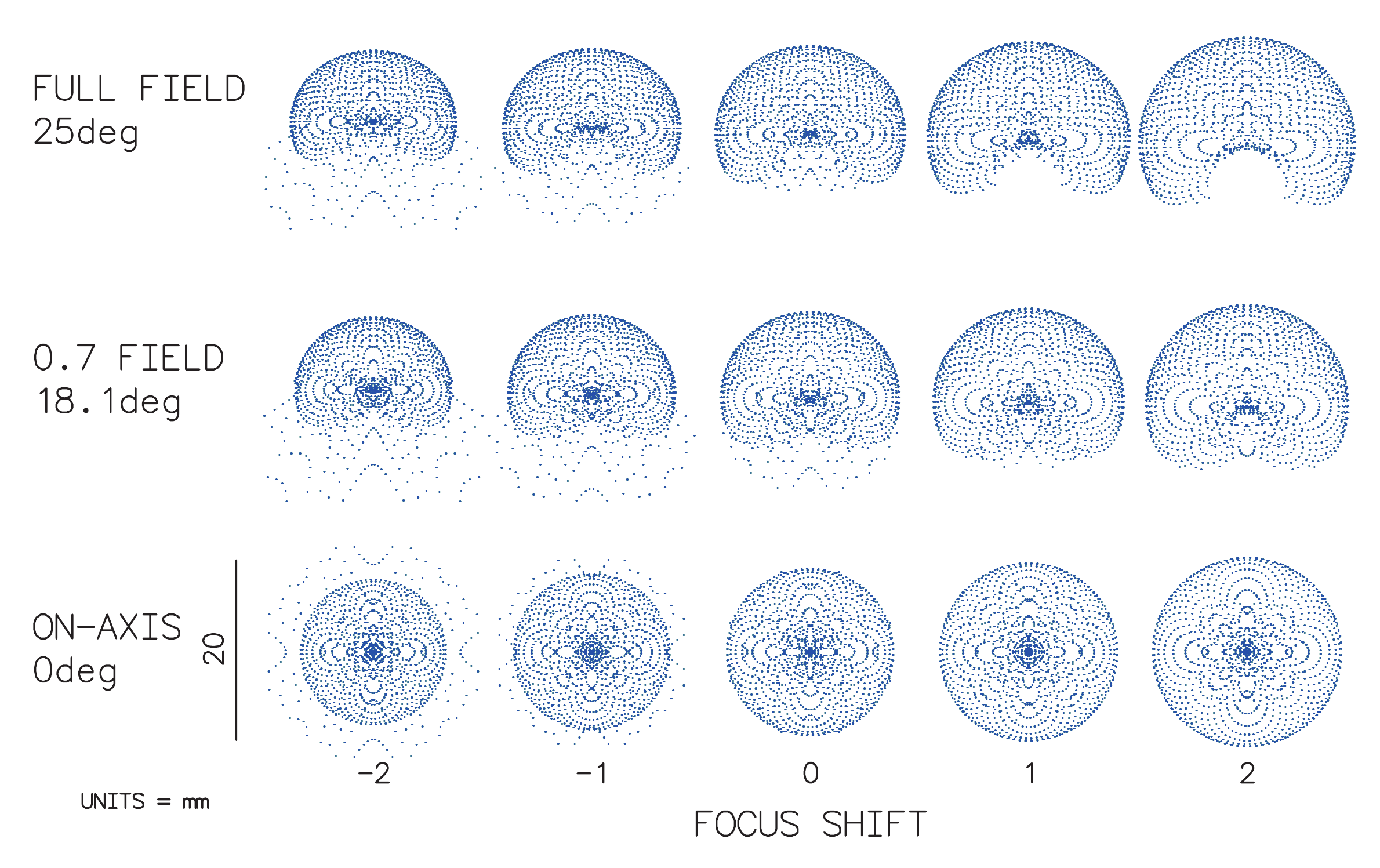}
\caption{Images of light spots on the focal surface of the detector from parallel light beams taking into account the shadow from the mosaic. All values are given in mm.}
\label{fig:spots}
\end{figure}

On the fig.\ref{fig:optic_sphere3} (right) the relative amount of light collected within the certain radius is shown. Different colors shows different incidence angle relative to the detector optical axis.
On the fig.\ref{fig:spots} the light spots on the photosensitive detector surface are show for different conditions - three light incidence angles on the detector (0$^\circ$, 18.1$^\circ$ and 25$^\circ$) and five offsets of the photosensitive surface. For the reference a 20~mm scale line is shown in the lower left corner. The spots were calculated for the 420~nm wavelength.

The mirror for the detector is planned to be constructed on the basis of composite materials (the cellular aluminium for the base). This design has sufficient rigidity and low weight.

\begin{figure}[t]
\centering 
\includegraphics[width=.55\textwidth]{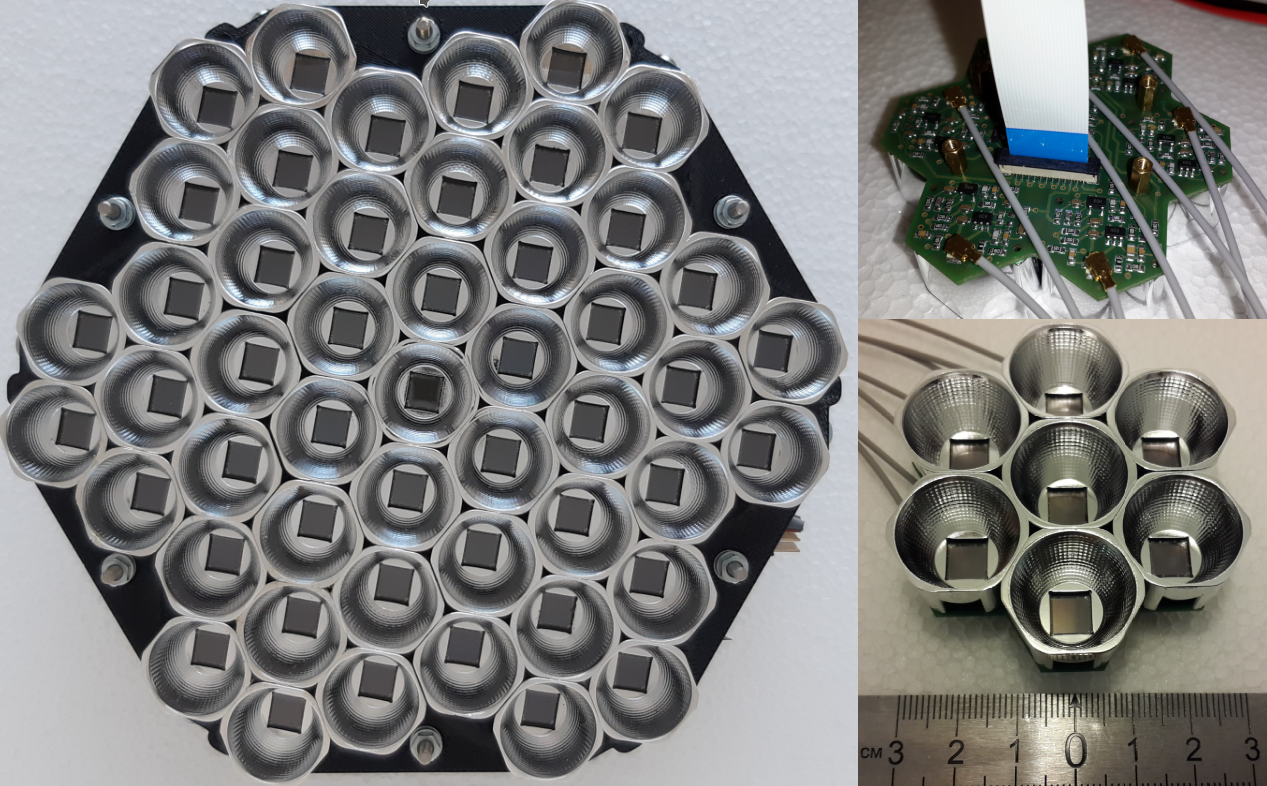}
\caption{The matrix prototype of 49 SiPM is assembled from seven electronic boards of 7 SiPM with preamplifiers.}
\label{fig:mosaic49_7}
\end{figure}

The main element of the new installation will be a segment of seven SiPM Micro FC-60035 SiPMs. The tests of a matrix of seven such segments (49 SiPM) was successfully completed (see fig.\ref{fig:mosaic49_7}). Each segment was equipped with seven preamplifiers and a temperature sensor to account for the effects of thermal emission and gain shift. Each SiPM was equipped with a CA10929 Boom-MC-W light collector with angular characteristic of $\pm$24 degrees at 50\% effectiveness. The assembled SiPM matrix was successfully tested as the sensitive element of the Small Imaging Telescope withing the TAIGA Project (for details see~\cite{SIT20}). In this project, it is planned to modify and adapt the SiPM segment for use in a ultra-wide angle optical system.

To register analogue signals from SiPMs, a digitization board based on the AnalogDevices ADS5296A 8-channel fast analog-to-digital converter (FADC) chip~\cite{FADC} will be developed. The sampling frequency of this chip is up to 80 MS/s (12.5~ns step) at 12-bit resolution and 100~MS/s (10~ns step) at 10-bit resolution.
The board design allows to reduce time of digitization by a factor of 2 by installing two FADC on each SiPM.
Small size of the chip (9x9 mm) allows significant reduction of the detector weight and dimensions. 
Digitized signals from each channel in serial code are transmitted to the LVDS interface on a XILINX Zynq FPGA module.
These modules are equipped with a built-in computer running the Linux operating system.
All internal logic of the measuring system and the trigger system is recorded in the chip as a configuration file (a program in the VHDL language of integrated circuit equipment description). This will allow more flexibility in the testing and in real measurements depending on the conditions.
The measurement results are recorded in the SSD drive of the on-board computer for further processing.

\section{Evolution of the experiment concept}

\begin{figure}[t]
\centering 
\includegraphics[height=.35\textheight]{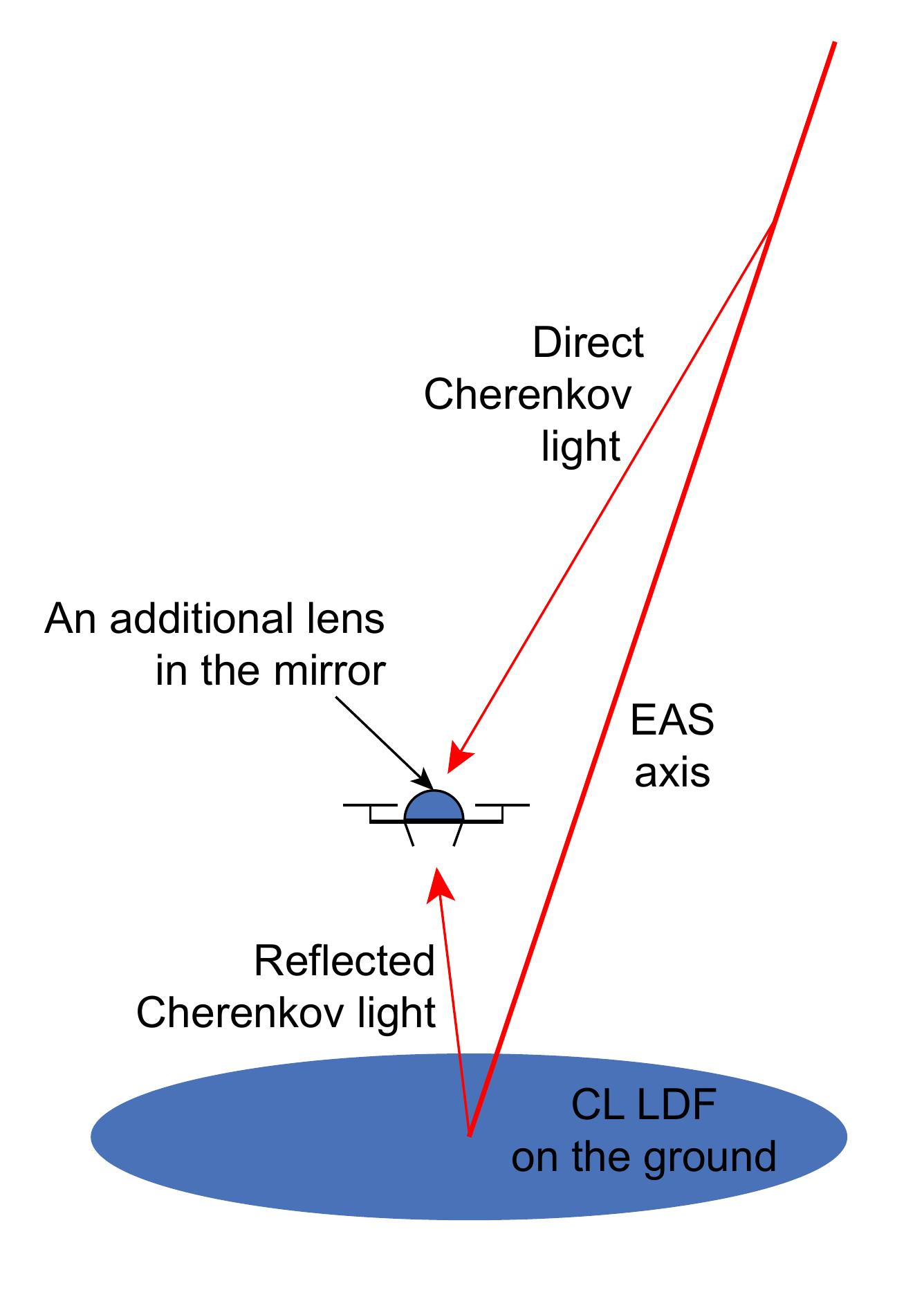}
\caption{Scheme of direct and reflected Cherenkov light of EAS.}
\label{fig:DirectCL}
\end{figure}

We generally follow the concept of SPHERE-2 experiment with one improvement and one major change. The former refers to the new camera ensuring more detailed analysis of the lateral distribution of EAS CL. The latter relates to the completely new information channel to be included in the SPHERE-3 detector.

The detector will use the simplified Schmidt optical system. In this system, the central part of the mirror is not used since it is in the shadow of the photodetector. A hole in the centre of the mirror with a wide-angle lens in it with an aperture of about 100 cm$^2$ will allow registration of the direct CL (see fig.\ref{fig:DirectCL}). Calculations show that for EAS 1 PeV particle the CL photons density is around 100 photons/cm$^2$ at the 100~m distance from the shower axis. Taking into account the SiPM quantum efficiency and losses of optical elements the total signal from the direct CL is expected at around 1000 photoelectrons.

The main function of the direct CL detection is to increase both the sensitivity and the specificity of the criterion for separation of actual EAS events using the correlated signals of direct and reflected CL. It is already clear that accurate detection and analysis of the CL angular distribution can yield substantial information on the PCR mass composition~\cite{Gal18a,Gal18b}. But such an experiment requires a large and more complex detector or even an array of such detectors placed wide apart, e.g. borne by a number of UAVs.

\begin{figure}[t]
\centering %
\includegraphics[height=.4\textheight]{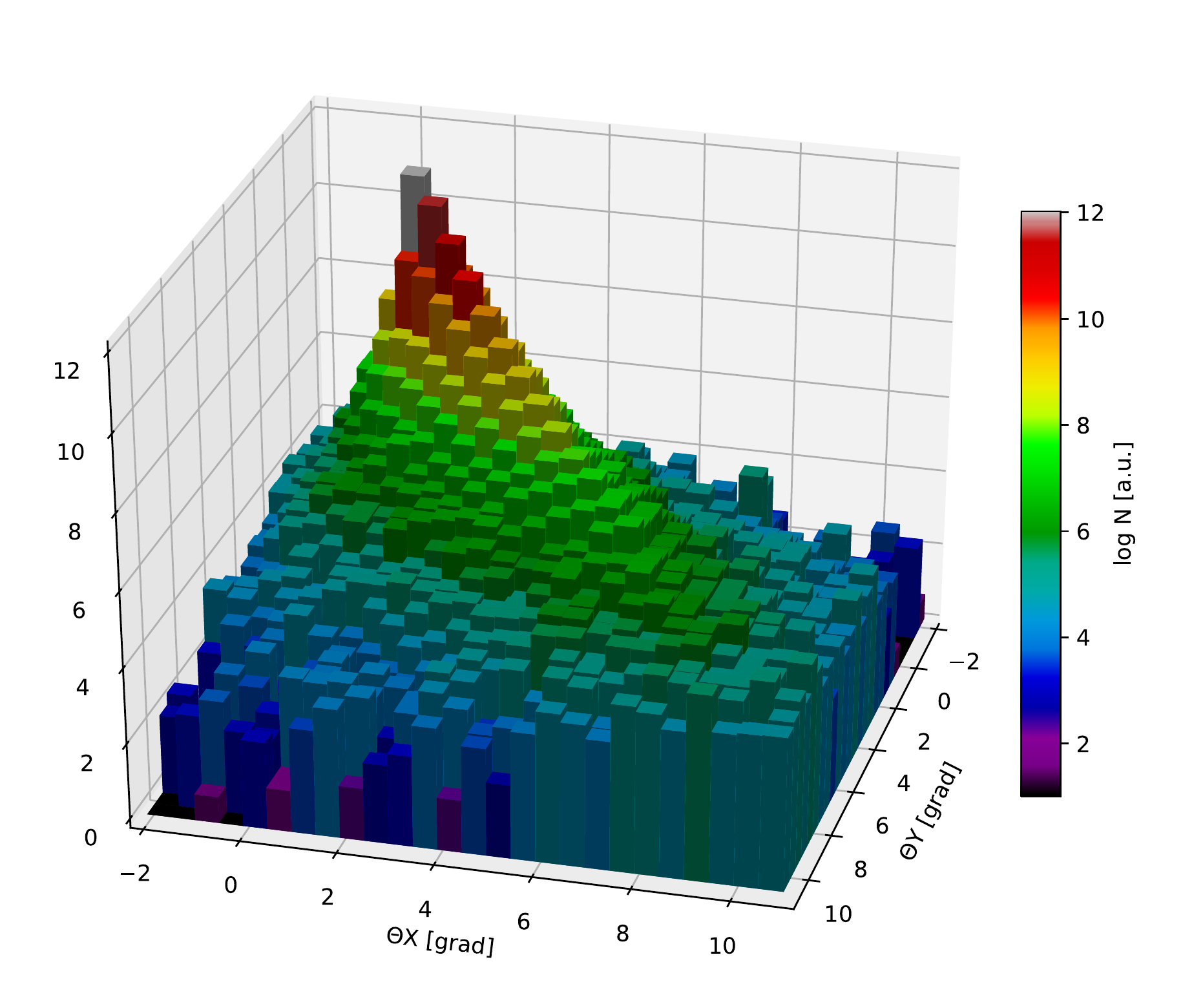}
\caption{Angular distribution of direct Cherenkov light from EAS. Primary particle: vertical 1 PeV proton. Core distance: 100 m. Observation level: 220 m above snowed surface.}
\label{fig:CLangular}
\end{figure}

Still, the second function of the direct CL detection channel is a preliminary study of the possibility to estimate the shape and angular size of CL flash (see fig.~\ref{fig:CLangular} for CL photons angular distribution example), which will help us to elaborate a new version of the detector capable of distinguishing between different primary masses using both lateral and angular CL characteristics.

For the moment the main means of primary mass separation are the criteria based on the shape of EAS CL lateral distribution over the snowed surface seen by SPHERE-3 optical system. Presumably such criteria will work better than with SPHERE-2 due to higher spacial resolution of the new detector.

Primary mass estimation is by far the most difficult part of the primary particle parameter evaluation problem. It is no wonder the PCR mass composition is still poorly known after more than half a century studies. Modeling of EAS CL characteritics~\cite{Gal18a,Gal18b,Ant09,Ant09b}
enables us to put forward some basic principles for choosing the criterial parameters for the primary mass separation:

\begin{itemize}
\item such parameter should be directly measurable or, at least, directly calculable from the measured quantities; this property may be called {\it observability};

\item another property is called {\it integrality}: the parameter must rely on the substantial part of the whole distribution measured (e.g., appreciable part of CL photons seen by detector); the property ensures the suppression of fluctuations which hinder the process of classification/separation;

\item one more important property of a criterial parameter is its {\it relativity}: it must reflect the shape of the distribution of the characteristic measured; this quality makes the parameter weakly dependent on the primary energy and, which is even more important, on the nuclear interaction model at super high energies; the latter feature has already been checked on EAS CL angular and lateral distributions~\cite{Gal18a,Gal18b,Ant09,Ant09b} but will likely hold on other shower characteristic distributions.

\end{itemize}
After all these points are taken into account one should optimize the definition of criteria parameter with respect to fluctuations. In other words, parameter values must have minimum possible variation for a given primary mass.


\section{The UAV as an experiment platform}

As a detector carrier at first stage it is planned to use  the DS1400~\cite{dronestroy} octocopter or a similar UAV.
The continuous operation time of this drone reaches 40 minutes.
The increase in exposure time is achieved by using several sets of charged batteries prepared before launch. 
To ensure the continuity of measurements implementation it is planned to use the additional UAV with the same detector (e.g. two detectors built).
At the project second implementation stage it is possible to use a group of several UAVs to proportionally increase the geometric factor and statistical reliability of the results. 

The use of hydrogen-air fuel cells~\cite{UAVair} or gasoline DELTA H1600H UAV~\cite{UAVoil} (or its analog) with a continuous operation time of up to several hours will greatly simplify the measurement process and improve the efficiency of experimental data collection.

To control the density and transparency of the atmosphere an auxiliary small quadrocopter DS550~\cite{dronestroy} UAV can be used with pressure, temperature, humidity sensors and a fast LED flash to produce artificial EAS-like light spots on the snow (in full spirit of the original idea~\cite{Chu74}).
The flash will be used to control the reflectivity of the snow and its geometry. Control of the atmosphere and reflection from the snow will improve the accuracy of measuring the EAS CL.

\section{Conclusion}

This project is the next realization of the reflected EAS CL registration method. The successful work with the SPHERE-2 detector gave a better understanding of this methods advantages and perspectives. The small scale of the detector allows to bypass many obstacles in the EAS registration such as calibration, timing and trigger conditions. The successful design and operation of the TAIGA Small Imagin Telescope shows that the SiPM matrix can be used as a sensitive element in telescope systems.

The proposed detectors optical system is also aimed at expanding the detection technique to accommodate direct CL measurements which will allow to combine two sources of EAS information in a single detector with a single sensitive element.

\acknowledgments
We warmly acknowledge the fundamental role in the development of reflected Cherenkov light registration method, pioneer works and important contributions to the project of our deceased colleague R.A.~Antonov.




\bibliography{Sphere3project.bib}




\end{document}